\documentstyle[aps,epsfig,amssymb]{revtex}

\def\la{{\langle}}
\def\ra{{\rangle}}

\begin{document}
\title{Fluctuation-response relation in turbulent systems}
\author{L. Biferale$^1$,
I. Daumont$^1$, G. Lacorata$^2$ and  A. Vulpiani$^{2}$}
\address{$^1$Dept. of  Physics  and INFM,
University of Rome "Tor Vergata",
Via della Ricerca Scientifica 1, I-00133 Roma, Italy}
\address{$^2$Dept. of  Physics and INFM, 
University of Rome "La Sapienza", P.le Aldo Moro 2, I-00185 Roma, Italy}

\maketitle
\begin{abstract}
\noindent
We address  the problem of measuring time-properties of Response
Functions (Green functions) in Gaussian   models (Orszag-McLaughin)
 and strongly non-Gaussian  models
 (shell models for turbulence). We introduce the concept
of {\it halving time statistics} to have a statistically
stable tool to quantify the time decay of Response Functions
and Generalized Response Functions of high order. We  show numerically
that in shell models for three dimensional turbulence Response Functions
are inertial range quantities. This is a strong indication that
the invariant measure describing
 the shell-velocity fluctuations is characterized
by short range interactions between neighboring shells.
\end{abstract}

\section{Introduction}

Fluctuation-Response (F/R) relation plays  an important role in
statistical mechanics and, more generally, in systems with chaotic
dynamics. With the term F/R relation one indicates the connection
between the relaxation properties of a system and its response to
an external perturbation. The relevance of this relation is 
evident: it allows to connect ``non-equilibrium'' features
(i.e. response and relaxation) to ``equilibrium''\cite{note1}
 properties
(correlation functions). As an important example, we mention the
Green-Kubo \cite{kubo85} formulas in the linear response theory which links
the response to an external field with correlations computed at
equilibrium.

Consider a system whose state is given by a finite dimension
vector ${\bf x}=(x_1,...,x_N)$, the average linear response
$G_i^j(t) \equiv \la  R_i^j(t) \ra$
is the average response after a time $t$ of the variable $x_i$ to
a small perturbation of the variable $x_j$ at time $t=0$.
Under rather general conditions (basically
one has to assume that the system is mixing) it is possible
to show that a generalized F/R relation holds
\cite{falcionietal90,carnevale91}:
\begin{equation}
\la R^i_j(t) \ra= \la {\delta x_i(t) \over \delta x_j(0)}\ra =
\la x_i(t) f_j[{\bf x}(0)] \ra,
\label{eq:FR}
\end{equation}
where the functions $f_j$ depend on the invariant probability
distribution $\rho({\bf x})$:
\begin{equation}
f_j[{\bf x}] = -{\partial \ln \rho({\bf x}) \over \partial x_j}.
\label{eq:rho}
\end{equation}
The physical meaning of (\ref{eq:FR}) is the following: consider
 a small perturbation 
$\delta {\bf x}(0)=(\delta x_1(0),...,\delta x_N(0))$ at $t=0$; the average
distance $\la \delta x_i(t) \ra$
from the unperturbed values $x_i(t)$ is
\begin{equation}
\la \delta x_i(t) \ra  = \sum_j \la R^i_j(t) \ra  \delta x_j(0).
\label{eq:pert}
\end{equation}
For Hamiltonian systems one realizes that (\ref{eq:FR}) is the usual
linear response theory. If $\rho({\bf x})$ is Gaussian, one has a simple relationship
between the response and the correlation function:
\begin{equation}
\la R^i_j(t) \ra =
{\la x_i(t) x_j(0) \ra - \la x_i\ra \la x_j \ra \over \la x_i x_j \ra - 
\la x_i\ra \la x_j \ra}.
\label{eq:gaussfr}
\end{equation}
In the general case, i.e. non-Gaussian statistics,  formula (\ref{eq:FR})
gives just a qualitative information, i.e. the existence of a link
between response and the general correlation function $\la x_i(t) f_j[{\bf x}(0)] \ra$.\\
\noindent
In particular, in the most interesting cases in which $\rho({\bf x})$
is unknown, it is extremely important that F/R relation (\ref{eq:FR})
exists, because it allows to control some properties of the
invariant measure, $\rho({\bf x})$,
in terms of the Response-Functions behavior. In the past, this
 has not always been clear,
e.g. some authors claim (with qualitative arguments) that in
fully developed turbulence there is no relation between equilibrium
fluctuations and relaxation to equilibrium \cite{rose78}, while a proper
statement would limit to the non existence of the usual ``Gaussian-like''
F/R relation (\ref{eq:gaussfr}). \\
\noindent
Response Functions have a clear phenomenological importance in many applied
problems where one needs to control and/or predict the system
reaction  as a function of external spatial and/or temporal perturbations.
Response functions, also known as Green functions,
 play also a very important role in many theoretical
attempts to attack non-equilibrium problems. In particular, in many
analytical approach to hydrodynamical
problems described by Navier-Stokes eqs, or models of them, Greens functions
naturally show up both in perturbative schemes \cite{lvov}
and  closure-like attempts like the DIA approximation \cite{kraichnan_dia}.

We stress that the F/R , in the form (\ref{eq:FR}),
 is a rather general relation
which does not depend too much on the details  of the
  measure of the systems, e.g.
both in presence or absence
of an energy flux.
For example,
in the field of  disordered systems the F/R had been  widely studied in
order to highlight
non trivial relaxation aspects e.g. aging  phenomena.
\noindent
In this paper we want to address the problem of F/R relation for the case
of dynamical models with many degrees of freedom and many characteristic
times. We are also interested in exploiting the F/R relation
 in models which exhibit strong departure
from Gaussian statistics. 
 We introduce a suitable new numerical
method for measuring the characteristic times involved in the response
functions.\\ This new method is based on the idea to characterize
the response behavior as a function of its {\it halving time statistics} (HTS),
 i.e. the
time, $\tau$,
 necessary for the response from a typical infinitesimal perturbation to reach,
say, one half of its initial value. \\
The plan of the paper is as follows. First
we investigate a dynamical system with many degrees of freedom and
many different characteristic times where still a classical Gaussian set of F/R
relations holds. The model is the so-called ``Orszag-McLaughlin'' model
which is used to probe the effective improvement
of halving-time-statistics with respect to the usual direct measurement
of time decaying properties.
 Than, we attack the much less trivial case of characterizing
response behavior in models for  three dimensional turbulent
energy cascade, i.e. shell models \cite{bjpv}.\\ Also in  the latter
case, the {\it halving time statistics} will allow us to measure
with good accuracy the non-trivial time properties
of Green functions. As a result,
we show  that also the response function (which probes  linear features
of the dynamical evolution)
 is  strongly affected  by the {\it non-linear} inertial range physics. 
As a consequence, 
short range interactions between neighboring shells are thought to characterize 
the invariant measure describing shell-velocity fluctuations.

\section{Numerical simulations}
Before entering the detailed description of the results, we want
to discuss a practical problem for the numerical computation of
$G^i_j(t) = \la R^i_j(t) \ra$.
In numerical simulations, $\la R^i_j(t) \ra$ is
computed perturbing the variable $x_i$ at time $t=t_1$ with
an ``infinitesimal''  kick of amplitude
$\delta x_j(t_1|t_1)
= \tilde{x}_j(t=t_1)-x_j(t=t_1)=\epsilon$, for $\epsilon \to 0$,
and the deviation $\delta {\bf x}(t|t_1) = {\bf \tilde{x}}(t)-{\bf x}(t)$
is computed integrating the two trajectories ${\bf x}$ and ${\bf \tilde{x}}$
up to a prescribed time $t_2=t_1+\Delta t$. At time $t=t_2$ the variable
$x_i$ is again perturbed with another kick, and a new
 sample $\delta {\bf x}(t)$ is computed and so forth.
The procedure is repeated $M \ll 1$ times and the mean response is then
evaluated as
\begin{equation}
G_i^j(t)  = {1 \over M} \sum_{k=1}^M {\delta x_i(t_k+t|t_k)
\over \delta x_j(t_k|t_k)}.
\label{eq:Rij}
\end{equation}
In presence of chaos, the absolute value of the deviation,
 $|\delta x_i(t_k+t,|t_k)|$, typically grows
exponentially with $t$. Therefore,  the mean response,
$\la R^i_j(t) \ra$,
is the result of a delicate
balance of terms with non-fixed sign. As a result, we
 have that the error $\sigma(t)$
on $\la R^i_j(t) \ra$ increases exponentially with $t$
\begin{equation}
\sigma(t) \sim {\exp(\gamma t) \over \sqrt{M}}
\label{eq:sigmatau}
\end{equation}
where $\gamma$ is the generalized Lyapunov exponent of second order
(greater or equal the maximum Lyapunov exponent).
 One  easily understands the main problem
in trying to numerically compute any response function for large times:
one needs to control an observable which is rapidly
decaying to zero with exponentially-large fluctuations.
In practice, it turns out to be impossible to have a reliable control on the asymptotic behavior 
of  Green functions (see next
sections and figures therein). \\
In order to avoid this trouble we propose another approach. 
Let us first consider only {\it diagonal} responses, i.e. response
after a time $t$ of the $n$-th variable from a perturbation of
the same $n$-th variable at time $t=0$,
$G_n^n(t) \equiv \la R^n_n(t) \ra$.

In this case, we claim that it is possible to have
a good characterization of the main temporal properties by looking
at the {\it halving time statistics} (HTS), that is at the probability density
functions, ${\cal P}(\tau)$  of the time $\tau$ necessary
to see an appreciable decay of the response
 function: $R^n_n(\tau=t)=\lambda R^n_n(0)$, with
the threshold $\lambda$ fixed to a macroscopic value,
 say $\lambda=1/2$. In practice,
one  performs many response experiments,
by collecting the statistics of the times necessary to
see the response  become one half of its initial value.
The advantage of this HTS with respect to the more standard way
of characterizing the mean response $G_n^n(t)$ with some
typical time is that one does not need to know any functional
behavior for the averaged response and, moreover, one has also a
control on the fluctuations of the characteristic times, i.e. the HTS integrates all times 
corresponding to {\it halving events}.  
In the following,
we  show that the HTS is at least able to reproduce with good
accuracy the same results of the direct fitting procedure of the
averaged response in  cases when the classical F/R relation (\ref{eq:gaussfr})
 holds
(Orszag-McLaughlin model, i.e. Gaussian statistics)
and, more interesting, it is also able to give new hints on the
F/R relations  when time-intermittency and strong departure from
Gaussianity are present (shell models). In the following we will also
discuss the cases of non-diagonal responses, $\la R^n_m(t) \ra$,
with $n \neq m$ and the cases of generalized higher order responses
$$\la R^{n_1,n_2,\dots,n_r}_{m_1,m_2,\dots,m_r}(t_1,t_1';t_2,t_2';\dots;t_r,t_r')\ra=
\la {\delta x_{n_1}(t_1) \over \delta x_{m_1}(t_1')}
 {\delta x_{n_2}(t_2) \over \delta x_{m_2}(t_2')} \dots
 {\delta x_{n_r}(t_r) \over \delta x_{m_r}(t_r')} \ra  $$.

\subsection{The Orszag-McLaughlin model}

Let us consider the following model \cite{orszag}:
\begin{equation}
{dx_n \over dt}= x_{n+1} x_{n+2} + x_{n-1} x_{n-2} - 2 x_{n+1} x_{n-1},
\label{eq:orszag}
\end{equation}
with $n=(1,2,...,N)$, $N=20$, and the periodic condition $x_{n+N}=x_n$.
This model contains some of the main features of inviscid hydrodynamics:
a) there are quadratic interactions; b) a quadratic invariant exists
($E=\sum_{n=1}^{N} x_n^2$);
c) the Liouville theorem holds.
For sufficiently large $N$
the distribution of each variable $x_n$ is Gaussian. In this situation,
classical F/R relationship exists for each of the $n$ variables: 
self-response functions to infinitesimal perturbations
are indistinguishable from the corresponding self-correlation functions
\cite{falcionietal90}.

We have slightly modified the system (\ref{eq:orszag}) in order to
have variables with different characteristic times. This can be done, for
instance, by rescaling the evolution time of each variable:
\begin{equation}
\frac{dx_n}{dt} = k_n  (x_{n+1} x_{n+2} + x_{n-1} x_{n-2} - 2 x_{n+1} x_{n-1}),
\label{eq:orszagdiff}
\end{equation}
where the factor $k_n$ is a function of the "number of identification"
(e.g. site in the chain)
of the variables defined as $k_n = \alpha \cdot \beta^n$, with 
$\alpha = 5 \cdot 10^{-3}$
and $\beta= 1.7$,
for $n=1,2,...,N/2$, with the "mirror" property $k_{n+N/2}=k_{N/2+1-n}$.
An immediate consequence is that the quadratic observable
 $E$ is no longer invariant during the time evolution of the system
(\ref{eq:orszagdiff}). The mean energy per mode, $E_n=\la x_n^2 \ra$
 (not shown), follows a linear law with $k=k_n$.
It can be demonstrated that a new quadratic integral of motion exists,
and this has the form:
\begin{equation}
I = \sum_{n=1}^{N} { x_n^2 \over k_n}.
\label{eq:int}
\end{equation}
Moreover, the  $x_n$ variables are shown
 to preserve the Gaussian statistics
to a good extent. Therefore, the only effect of the change in the original
Orszag-McLaughlin system is that each variable now has its own characteristic
time.

\begin{figure}
\includegraphics[angle=-90, width=0.75\textwidth]{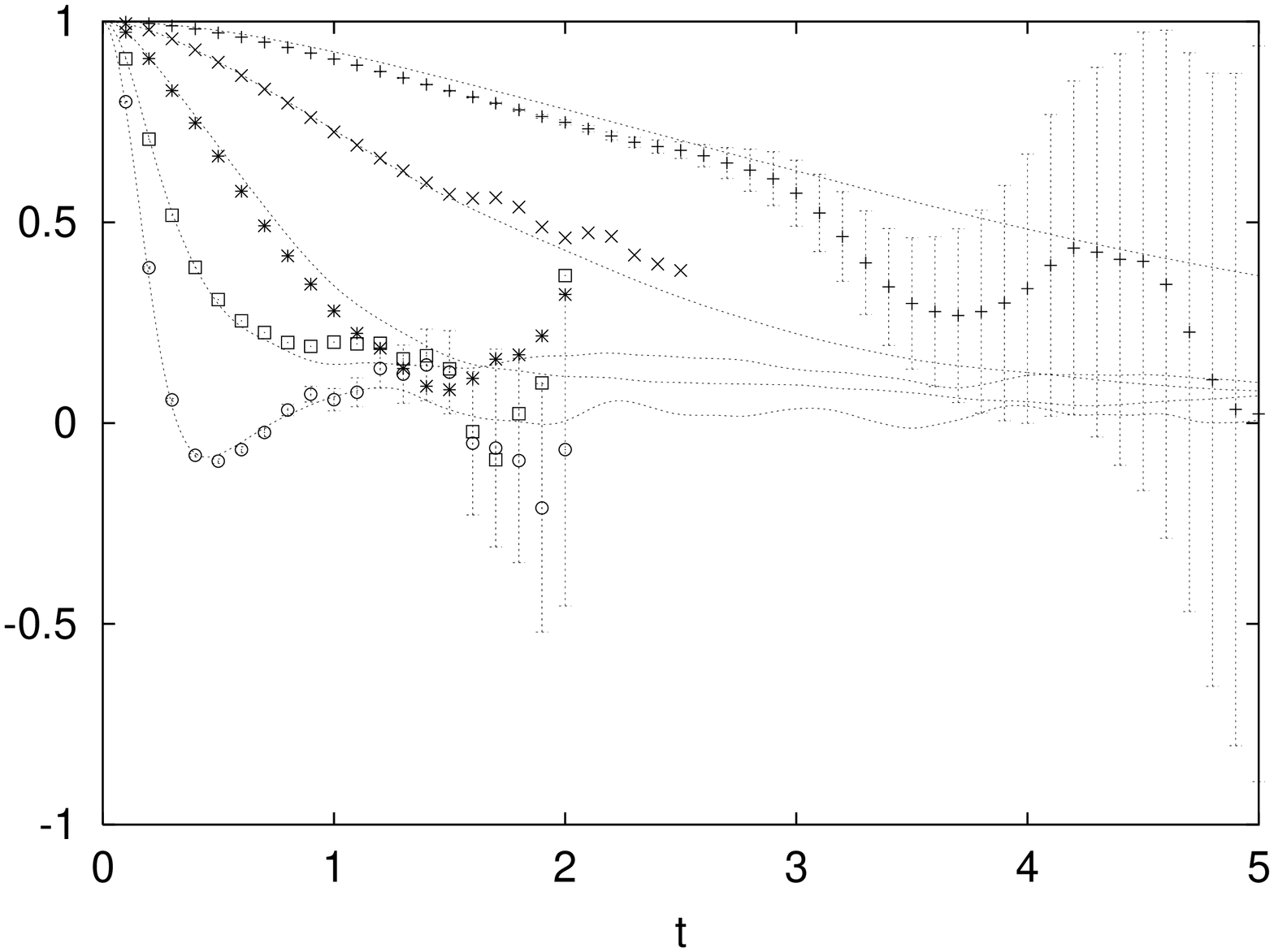}
\caption{Plot of the averaged Response Functions $G_n^n(t)$ and 
Correlation Functions $C_{n,n}(t)$ for five fast  variables of 
the modified Orszag-McLaughlin model, 
$n=6$ ($+$), $n=7$ ($\times$), $n=8$ ($\ast$), $n=9$ ($\square$) 
and $n=10$ ($\circ$). 
Statistical error bars are shown only for Response Functions  
corresponding to $n=6$ and $n=10$.  
Thin lines represent Correlation Functions. 
The statistics is over $10^5$ events.  
}
\label{fig:flures}
\end{figure}

Let us see how correlation and response functions behave for the
system (\ref{eq:orszagdiff}). In Figure~\ref{fig:flures} the self-correlation
functions
\begin{equation}
C_{n,n}(t)={\la x_n(t)x_n(0)\ra-\la x_n\ra^2 \over
\la x_n^2 \ra - \la x_n\ra^2 },
\label{eq:selfcorr}
\end{equation}
and the self-response functions
\begin{equation}
R_n^n(t)=\la{\delta x_n(t) \over \delta x_n(0)}\ra,
\label{eq:selfresp}
\end{equation}
 are shown. As a consequence of the preserved
Gaussian statistics, F/R relation of the form (\ref{eq:gaussfr}) holds 
for each of the variables, at least over time delays not too long.
The (linear) response functions are computed as decay functions of
single variable perturbations to infinitesimal
instantaneous "kicks", averaged over a large number of simulations.
If we conventionally define the correlation time of a
variable $x_n$ as the time delay $\tau_{C}(n)$
after which the correlation function
becomes lower than the value $1/2$, 
we find that:
\begin{equation}
\tau_{C}(n) \sim k_n^{-3/2}
\label{eq:CT}
\end{equation}
The exponent of the scaling law (\ref{eq:CT}) can be explained with a 
dimensional argument, by noticing that from the mean energy per variable 
we get $x_n^2 \sim k_n$, so from (\ref{eq:orszagdiff}) and 
(\ref{eq:int}) the characteristic 
time results to be $\tau_C(n) \sim k_n^{-3/2}$. We notice that the scaling 
(\ref{eq:CT}) is robust with respect to the choice of the threshold value, 
$\lambda$, i.e. it is  observed 
even if the decay factor $\lambda$ is chosen slightly different from $1/2$.   
 
The response time
 $\tau_R(n)$ is
 defined as the time interval after which the averaged response function
becomes lower than $1/2$. 
 We must observe that the computation of 
the mean response function is practically impossible after a certain 
time delay, because of exponentially growing errors. 

\begin{figure}
\includegraphics[angle=-90, width=0.75\textwidth]{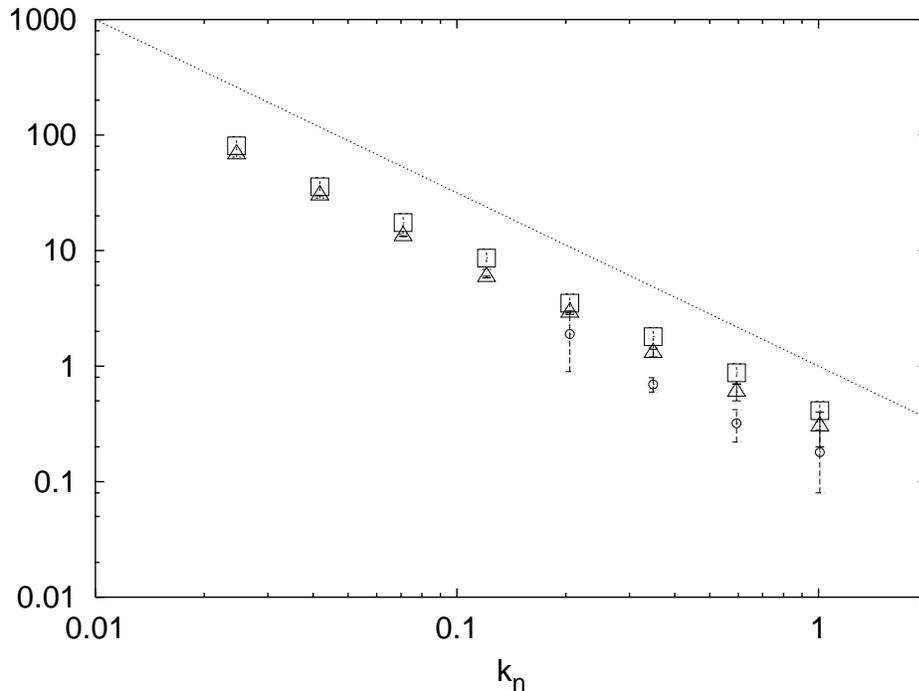}
\caption{
Log-log plot of Correlation Times $\tau_C(n)$ ($\triangle$), 
Mean Halving Times $\langle \tau(n) \rangle$ ($\square$), 
and Response Times  
$\tau_R(n)$ ($\circ$) 
as function of $k_n$, for the modified  
Orszag-McLaughlin model. Notice the much larger errors
found when measuring the  characteristic Response Times, $\tau_R(n)$. 
Errors on $\tau_C(n)$ and $\langle \tau(n) \rangle$ are of the same size 
as the representative symbols.  
The statistics 
is over $10^5$ realizations. 
All these characteristic times follow the same scaling law with 
$k_n$. The 
exponent $-3/2$ of the scaling law follows from dimensional arguments. 
}
\label{fig:HTHRCT}
\end{figure}

Last,  halving times $\tau(n)$
 have been computed always for the
variables of the system (\ref{eq:orszagdiff}), 
 using the same procedure as before (i.e. infinitesimal kicks).
 The halving time PDF's decay
exponentially and, as can be shown,
they all can be ``collapsed'' to the same renormalized PDF
for a proper
rescaling of the halving time (see below).
A comparative plot of $\la \tau(n) \ra$,  
$\tau_C(n)$ and $\tau_R(n)$ 
is shown in Figure~\ref{fig:HTHRCT}. 
Halving times and correlation times follow the same scaling law 
with $k_n$ and, for each variable, have values very close to each other. 
Typical time decaying of the averaged Response, $\tau_R(n)$ 
are very   difficult to estimate due to high errors for slow variables.
 In fact, 
only a few points are shown in Figure~\ref{fig:HTHRCT}, the ones for which 
the mean response drops down to $1/2$ fast enough, before the statistical 
error  
become too large (say larger than $100\%$).  
 The advantage of the HTS with respect to the mean response function 
is that, with the same statistics, 
halving times can be computed for all variables within reasonable uncertainty, 
while response times, $\tau_R(n)$, 
 are generally affected by exponentially growing 
errors and are practically not defined when the typical relaxation time scale 
is longer than the error growth time scale.

\begin{figure}
\includegraphics[angle=-90, width=0.75\textwidth]{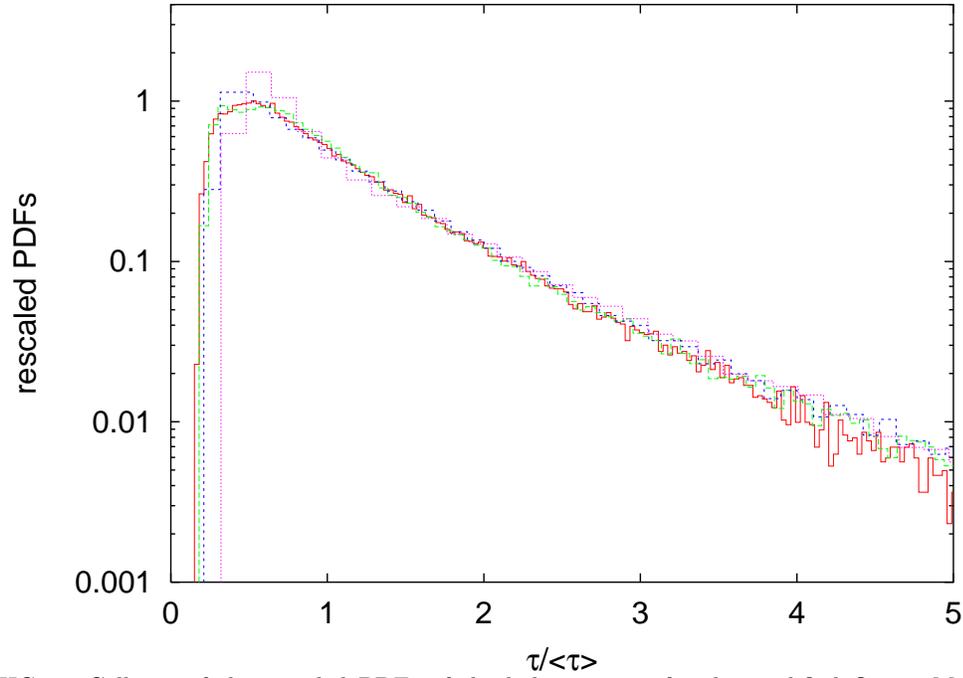}
\caption{Collapse of the rescaled PDFs of the halving times 
for the modified Orszag-McLaughlin model. 
For simplicity, only
 the PDFs relative to the fastest four variables are shown, 
$n=7,...,10$. 
The statistics is over $10^5$ impulsive 
infinitesimal perturbations as in Figure~\ref{fig:HTHRCT}. 
}
\label{fig:orszagpdf}
\end{figure}

The numerically computed PDF's of the halving time $\tau$ can be rescaled 
as follows:
\begin{equation}
\tau \to {\tau \over \la \tau \ra}, \;\;\;\;P(\tau) \to \la \tau \ra P(\tau).  
\label{eq:rescaling}
\end{equation}
We show in Figure~\ref{fig:orszagpdf} the overlap of some rescaled PDF's 
of the halving times. 
As a consequence, all moments of the $\tau(n)$ PDF's have a simple scaling 
$$
\la \tau(n)^p \ra \sim k_n^{-{3 \over 2}\cdot p}.
$$

We have found that in a Gaussian case, the
response to infinitesimal perturbations can be characterized both
with the classic mean response function and with the mean halving time
technique. It is worth stressing that the HTS  could be
the only technique usable for studying relaxation to non linear
perturbation in complex systems.

\subsection{Shell model}
Shell models for turbulent energy cascade
have proved to share many statistical properties
with  turbulent three dimensional velocity fields
\cite{frisch,bjpv}. Let us
introduces  a set of wavenumber $k_n = 2^n k_0$ with
$n = 0,\dots,N$.
The shell-velocity variables $u_n(t)$ must be understood as
the velocity fluctuation over a distances $l_n=k_n^{-1}$.
It is possible to write down many different sets
of  coupled  ODEs possessing the same kinematical features necessary
to mimics Navier-Stokes non-linear evolution.
In the following we will present numerical
results for a particular choice, the so-called Sabra model
\cite{sabra}, namely:
\begin{equation}
\left(\frac{d}{dt}+\nu k_n^2\right) u_n =
i\left[k_nu_{n+1}^*u_{n+2} + b k_{n-1}
u_{n+1}u_{n-1}^* + (1+b)k_{n-2}u_{n-2}u_{n-1}\right] +f_n,
\label{shell_v}
\end{equation}
where $b$ is a free parameter, $\nu$ is the molecular viscosity and
$f_n$ is an external forcing acting only at large scales, necessary
to maintain a stationary temporal evolution. The main,
strong, difference with the model discussed in the previous section
consists in the existence of a mean energy flux from large to
small scales which drives the system  toward a strongly non-Gaussian
stationary temporal evolution \cite{pissa}.
 Shell models here discussed presents exactly the same qualitative
difficulties of the original Navier-Stokes eqs: strong non-linearity
and far from equilibrium statistical fluctuations. The most striking
quantitative feature of the non-Gaussian  statistics is
summarized in the existence of anomalous scaling laws of velocity
moments:
\begin{equation}
\la |u_n|^p \ra \sim k_n^{-\zeta(p)},
\end{equation}
with $\zeta(p) \neq p/2 \zeta(2)$. Anomalous scaling, also known
 as intermittency, is the quantitative way to state
that velocity PDF's at different scales cannot be rescaled
by any  changing of variables.\\

\begin{figure}
%\centerline{\psfig{file=fig1.tex.eps,width=6cm}}
\includegraphics[width=0.75\textwidth]{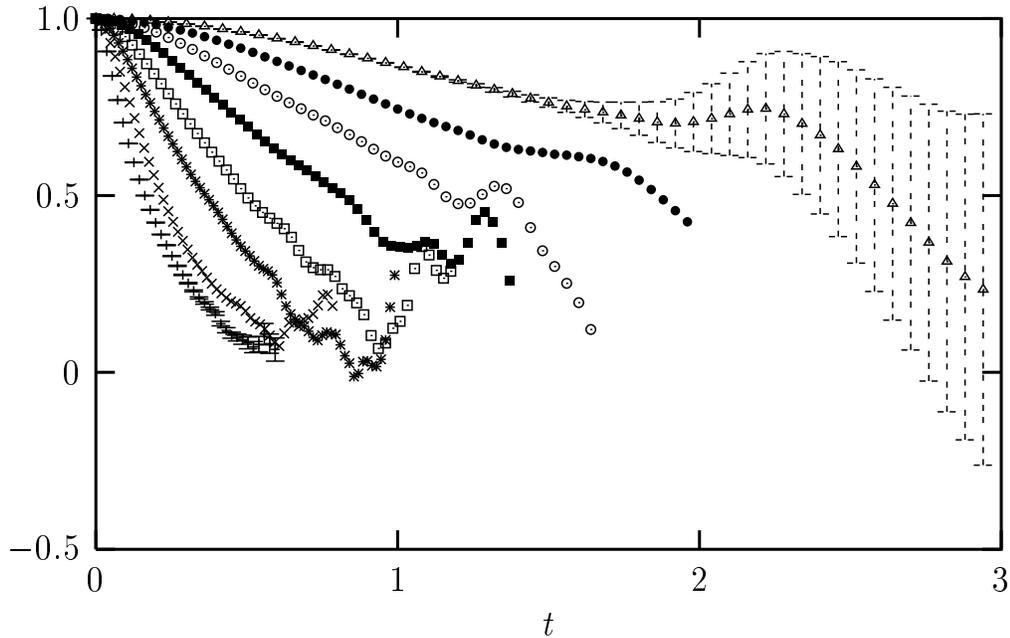}
\caption{
Modulus of the average response functions, $G_n^n(t) = \la R_n^n(t) \ra$,
 for  shells $n=7,\dots, 14$ (from top  to bottom).
Error bars are shown only for the smallest and the largest scales.
The number of independent kicks used to perform the averages
 is around $2.10^5$. Notice the extremely large error bars
measured for the slowest shell variables.
The parameters entering in the eqs. of motion  (\ref{shell_v}) are $b=0.4$,
$\nu=5.10^{-7}$, for $N=25$ shells.}
\label{RF7to14}
\end{figure}

\begin{figure}
%\centerline{\psfig{file=RepCor10.tex.eps,width=6cm}}
\includegraphics[width=0.75\textwidth]{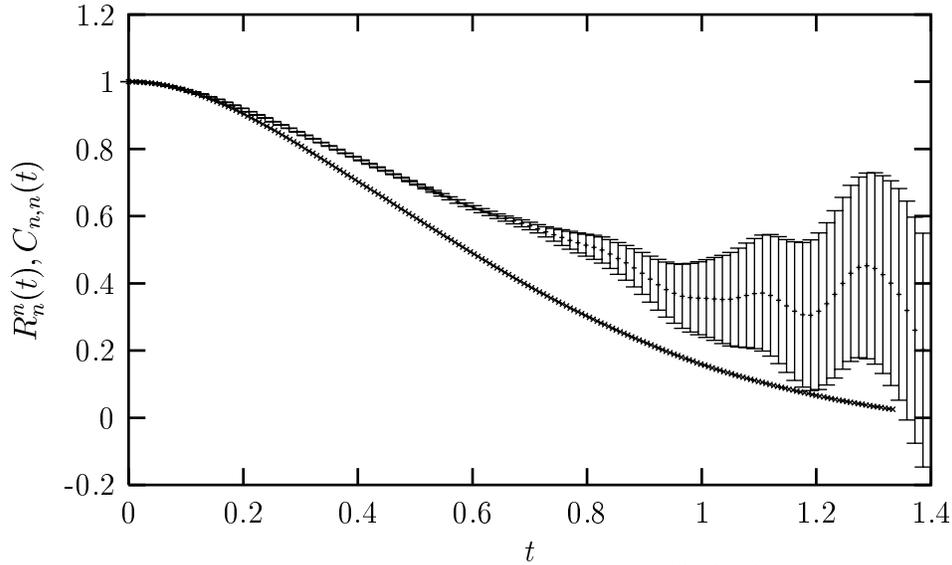}
\caption{Comparison
 between the averaged Response Function, $G_n^n(t)$, (top)
 and the self-correlation, $C_{n,n}(t)$ (bottom)
for the shell $n=10$. Notice the different order of magnitude of error bars.}
\label{RepCor10}
\end{figure}

\begin{figure}
%\centerline{\psfig{file=traj.tex.eps,width=6cm}}
\includegraphics[width=0.75\textwidth]{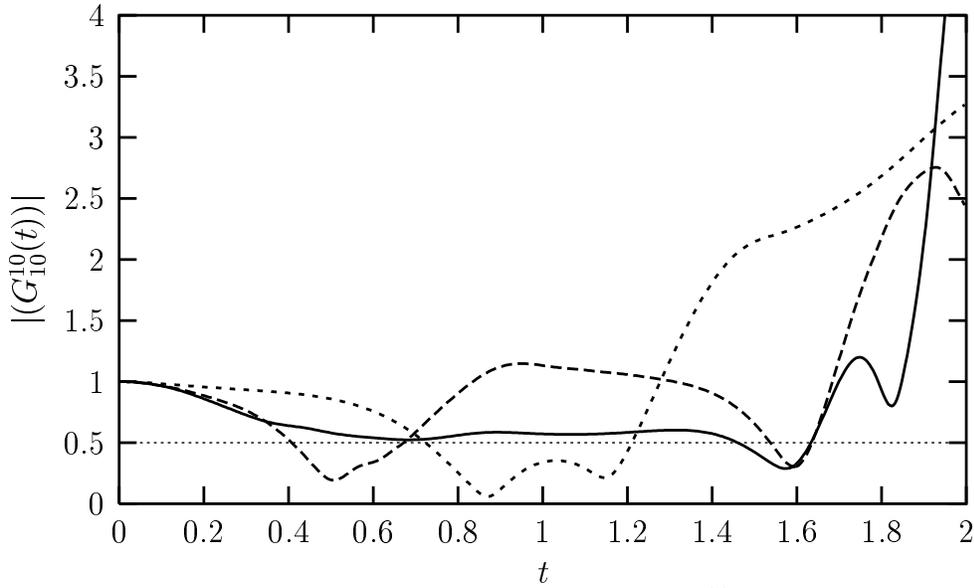}
\caption{Plot of three different diagonal instantaneous responses, $R_n^n(t)$,
 for the shell  $n=10$, versus time. Halving time is fixed by the first
time when the curve touches the threshold at $\lambda=1/2$. }
\label{3traj}
\end{figure}

Let us now discuss two subtle points. Using some general arguments
from the dynamical systems theory, one has that all the (typical)
correlation
functions at large time delay have to relax to zero with the same
characteristic time, related to spectral properties of the
Perron-Frobenius
operator. If one uses this argument in a blind way, the apparently
paradoxical result is that all correlation functions,
 $C_{n,n}(t)=\la u_n(t)u_n(0) \ra$, must go
to zero with same characteristic times.
On the contrary one expects a whole hierarchy of characteristic
times distinguishing the behavior of the correlation
functions at different scales \cite{phys_D}. In particular,
the self-correlation function,  $C_{n,n}(t)$, decays with a
characteristic time decreasing with $n$.
 The paradox is only apparent
since the dynamical systems argument is valid at very long times,
i.e. much longer than the longest characteristic time, and therefore
in systems with many different time-fluctuations it is not helpful.
In fact,
it is well established numerically and well understood theoretically
\cite{phys_D,bbt_prl,lvov_fr}
that general multi-scale multi-time correlation functions of the kind
$C_{n,m}^{p,q}(t) = \la |u_n(0)|^p |u_m(t)|^q \ra$
are  described by the cascade formalism, In particular,
most of the statistical properties in the inertial range
can be well parametrized by
the multi-fractal-formalism. \\
On the other hand,  the response properties are related to infinitesimal
perturbations. Therefore, a priori, it is not obvious that the response
depend on inertial range properties. The existence of the F/R
relation and the fact that $C_{n,n}(t)$
 (and other similar correlation functions)
are determined by the inertial range properties suggest  that also
the response features are ruled by the inertial range behavior if the
invariant measure is dominated by local interactions among shells. \\
Let us now examine the numerical results concerning Response functions
in the shell model. \\
Figure~\ref{RF7to14} shows the diagonal mean response,
 $G_n^n(t)$,  for a range of inertial shells, $n\in[7-14]$.
The most striking property is the impossibility to
follow the response behavior at large scales (small shells)
 for large times, i.e. the explicit evidence that errors grow
exponentially.
  In order to compare the different behavior (and different error
propagation) between response and correlation function, we plot
in Fig. (\ref{RepCor10}) both  the average response and the
self-correlation, $C_{n,n}(t)$, for the shell $n=10$.
 As it is clear from the previous figures, only
 response at the smallest scales (fast scales) in the inertial range
 can be computed with enough accuracy to follow an asymptotic
  decay. Still, also for this response the clear departure
between the response and the self-correlation shows is another
indication that the inertial-range statistics is far from  Gaussian.
By using HTS we can get an information on the temporal dependence
of response function in the whole range of inertial shells.
 In Figure~\ref{3traj},
we report some realizations of the instantaneous response
function, which shows typical halving time experiments.

\begin{figure}
%\centerline{\psfig{file=fig1.tex.eps,width=6cm}}
\includegraphics[width=0.75\textwidth]{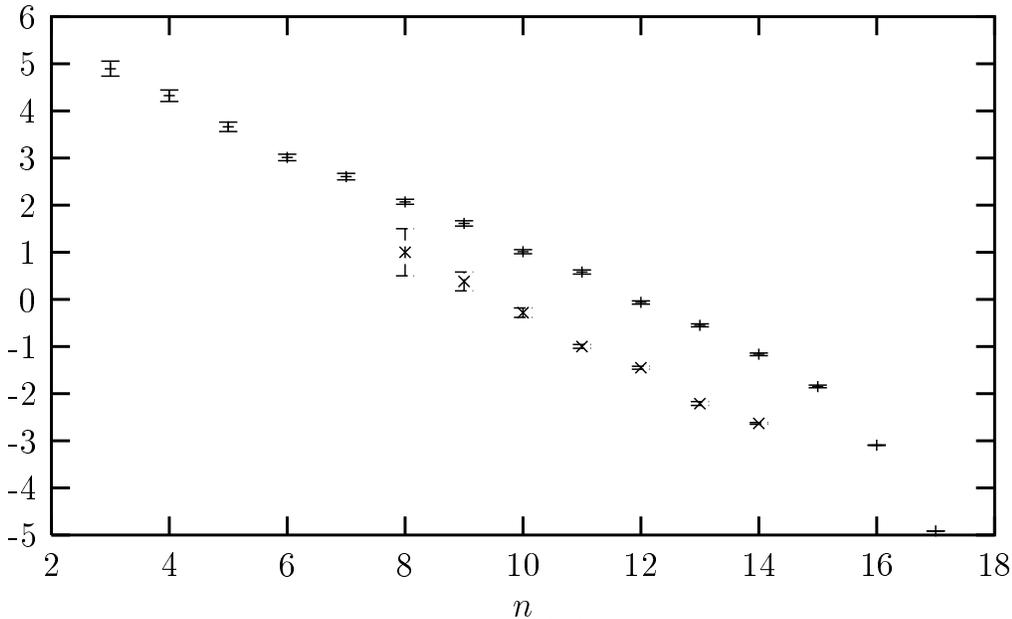}
\caption{Log-log plot of the {\it mean halving times},
$\langle \tau(n) \rangle$ ($+$) and
of decaying time of the mean diagonal response, $\tau_R(n)$ ($\times$), versus $k_n$.
We have checked that a different choice of the threshold $\lambda=1/2$ used to compute halving time does not affect the slope of the graph.}
\label{HTRTCTsm}
\end{figure}

In Figure~
(\ref{HTRTCTsm}), we summarize the results we obtain by comparing
the {\it mean halving time},  $\la \tau(n) \ra$, with the characteristic
times one extract from the decay properties of both mean response,
$\tau_R(n)$,
and correlation functions, $\tau_C(n)$,
 for those shells where such a behavior
can be safely extracted. It is worth noticing how the {\it mean halving
time}
allows a full characterization of time properties also for those
shells where the mean response $G_n^n(t)$ cannot be measured for large time
legs,  $t$.
Also, the dependence from the scale of the mean halving time
is given as a  best fit  $\la \tau(n) \ra
\sim k_n^{-\chi}$, with $\chi=0.53 \pm 0.03$.
 The value $\chi=0.53 \pm 0.03$ can be seen as an intermittent
correction to
the dimensional inertial-range prediction $2/3$.
On the other hand,  the dependency from the scale of $\tau_R(n)$ is
difficult
to extract due to the small number of points available.
%
% where we add the partial results about the
%time relaxation $\tau_{R}(n)$. It is worth to emphasize that
% the slope of this latest graph do not depend on the particular
%threshold $\lambda=1/2$ used for computing the halving times. It can
% be also observed that the exponents of the scaling law  of
%$\la \tau_{1/2}(n)\ra$ and $\tau_{R}(n)$ are slightly different,
%in contrast with the
%results of the Orszag-McLaughlin model. Because of intermittency,
% there is indeed no reason to expect an equality beetwen both.

\begin{figure}
%\centerline{\psfig{file=fig1.tex.eps,width=6cm}}
\includegraphics[width=0.75\textwidth]{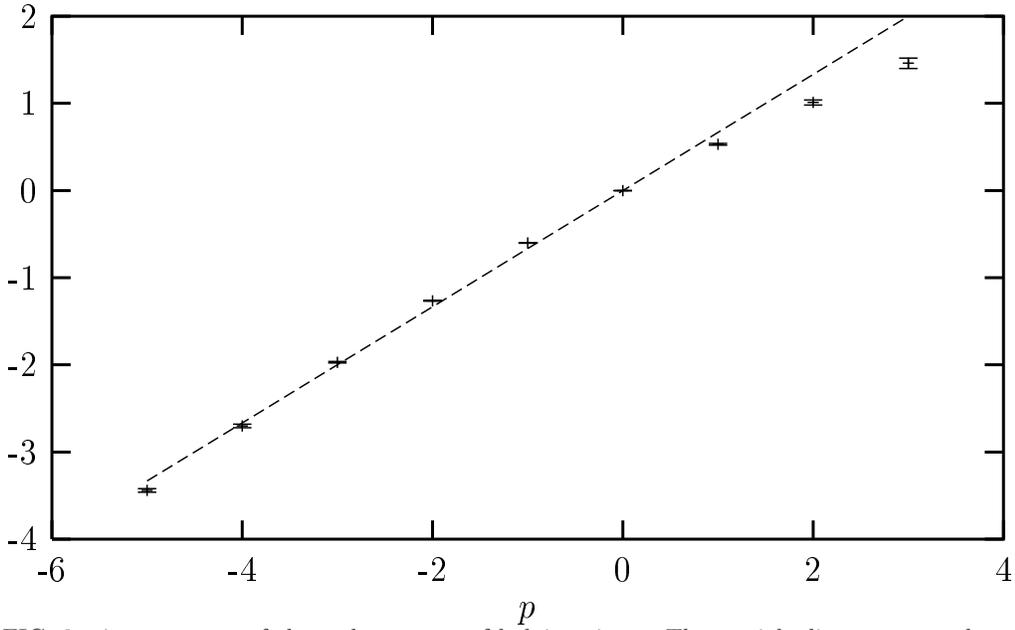}
\caption{$\psi_p$ exponents of the $p$-th moment of halving times.
The straight line corresponds to the dimensional prediction :
$\psi_p=2/3 p$.
}
\label{psip}
\end{figure}

Let us now focus on the whole PDF of the halving time statistics.
We first analyze the positive and negative moments of the halving times:

\begin{equation}
T^p(n) =  \la (\tau^p(n)\ra \sim  k_n^{-\psi(p)},
\end{equation}
with $p=-5,...,3$. Dimensional, non intermittent, scaling
would predict the linear behavior for the scaling exponents:
$\psi(p)=2/3 p $. In Figure~\ref{psip}
we plot the results for the halving times
scaling exponents $\psi(p)$ for
all moments from $p=-5,\dots,3$ and the straight line
corresponding to the dimensional inertial range  estimate.
 We notice that intermittent corrections
are much stronger for the
positive  moments than for the negative  moments. This must be related
to the
fact that positive moments of the halving times are dominated
by rare events where the response has a very long decaying.
We interpret the fact that $\psi(p) \simeq (2/3)p$ as an indication that  
linear Response Functions are inertial range quantities. The latter
results leads to the important conclusion that the invariant measure
is well approximated by short range interaction among shells
in the inertial range.

\begin{figure}
%\centerline{\psfig{file=HTnd.tex.eps,width=6cm}}
\includegraphics[width=0.75\textwidth]{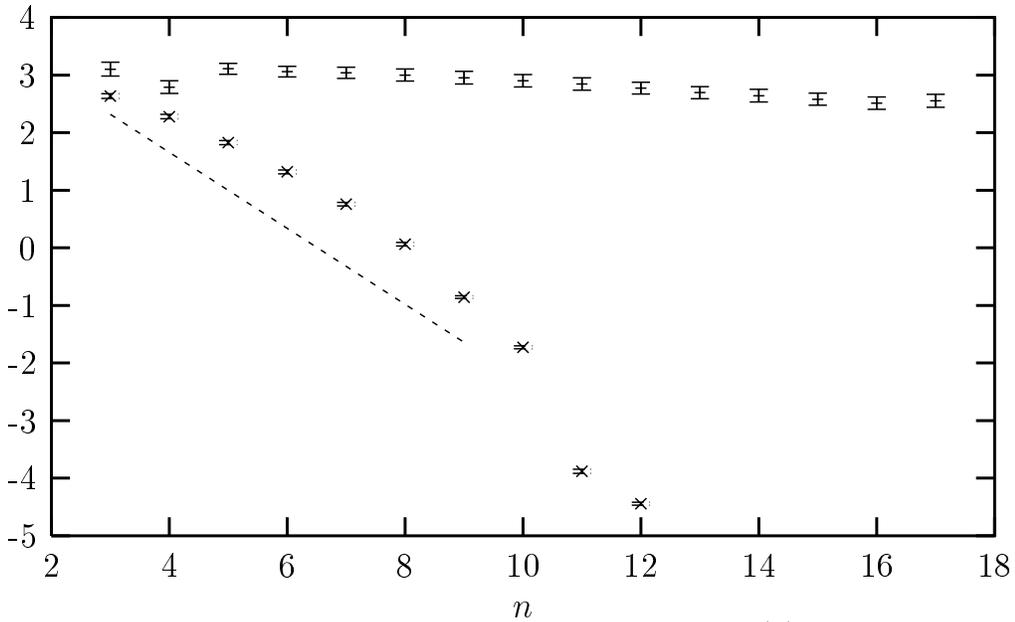}
\caption{Log-log plot of the off-diagonal response
characteristic times $\tau^{(m)}_n$ versus $k_n$.
We performed two experiments.
 First we perturb at large scales, $m=2$, and we follow
the response at small scales $n>2$,  ($+$); the expected independence
of characteristic times from the scale is well reproduced.
Second, we perturb at small scales, $m=13$, and we follow the response at
larger scales, $n<13$, $\times$. In the latter case, for comparison
we also plot the straight line (dashed)  with  the expected
dimensional slope $-2/3$.}
\label{HTnd}
\end{figure}

\begin{figure}
%\centerline{\psfig{file=grnl13.tex.eps,width=6cm}}
\includegraphics[width=0.75\textwidth]{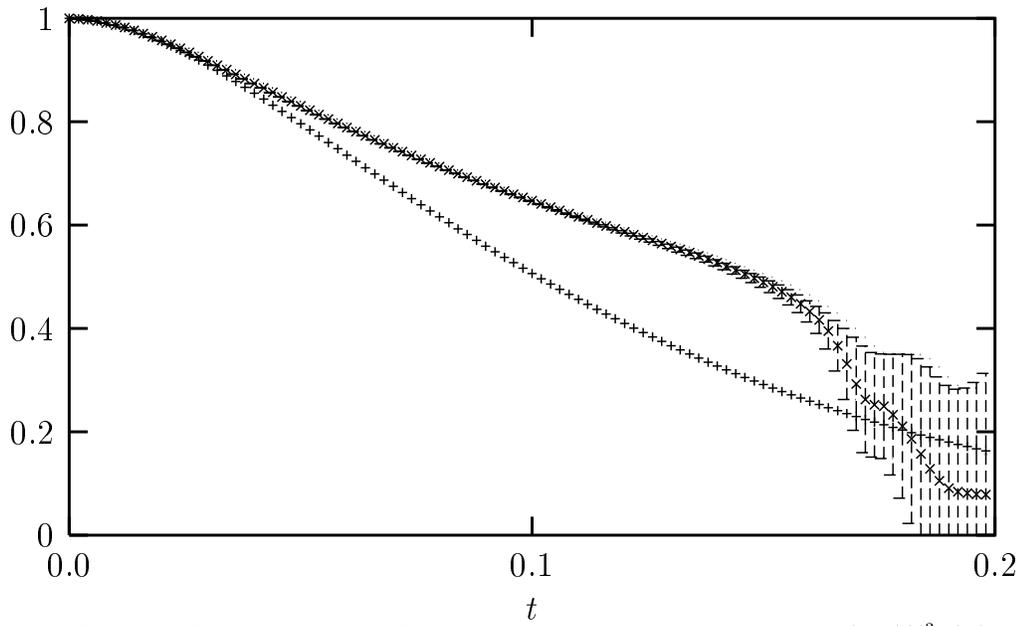}
\caption{Comparison between the third power of the mean diagonal
response, $|G_n^n(t)|^3$, ($+$) and  the generalized third order response,
$|S_n^n(t)|$, ($\times$) computed for
the shell $n=13$, with $5.10^5$ kicks.
}
\label{grnl13}
\end{figure}

As for the non-diagonal response function and for the generalized
response
function of higher order the numerical problems to measure them
are even more pronounced. First, let us examine the off-diagonal
response function $R_n^m(t) =
  \delta u_n(t)/ \delta u_m(0) $. Of course these
responses start from zero at time zero instead than from one as in
the diagonal case. Measure them by a direct average
is strictly impossible because of very large errors.  We still
decided to measure their characteristic time by using the time
the response reach a ``macroscopic'' fraction, say $1/2$,  of
the typical fluctuations on the scale where we are measuring the
response,
i.e. we collect the statistics of the first times $\tau$ such that
$R_n^m(t=\tau) = 1/2 \la |u_n|^2 \ra$.
 We expect a strong asymmetry of the characteristic
times depending whether the perturbation is done at smaller ($m > n$)
or larger ($n > m$) scales. Indeed, by using usual inertial
range arguments we expect that the response reacts always with
a typical time given by the time of the largest between
the two shells involved $n,m$. By fixing therefore the shell
where we perturb, say $m$, we expect that the typical time
of $R_n^m(t)$, $\tau_n^{(m)}$ is constant when $n>m$ and
scales as $k_n^{-2/3}$ when $n<m$. In Figure~\ref{HTnd}   we show that
indeed this behavior is well reproduced numerically.\\
The strong intermittency shown by halving times in Figure~\ref{psip}
is the clear signature of deviations from simple Gaussian-like
behavior of response functions. We must therefore also expect that
generalized responses of higher order are not simply
related to the linear response. For example, let us consider the third
moments
of the linear response\cite{note}:
$$
S_{n}^n(t)  \equiv \la (R_n^n(t))^3 \ra. $$
A simple non-intermittent behavior would suggest that $S_n^n(t) \propto
(G_n^n(t))^3$ while in  Figure~\ref{grnl13} it is possible to see that
this is definitely not the case for all $t$ legs where we have a
measurable signal. Unfortunately the already discussed statistical
problems in measuring averaged response functions for long times
are even more pronounced for generalized response functions. Therefore
we refrain from showing any results for the scaling behavior
of typical times of the generalized response functions. \\

\section{Conclusions}
We have addressed the problem of measuring time-properties of response
functions in Gaussian   models (Orszag-McLaughin)
 and strongly non-Gaussian models
like shell models for turbulence. We have introduced the concept
of {\it halving time statistics} with the aim to have a statistically
stable tool to quantify the time decaying of response functions
and generalized response functions of high-order. We have shown
numerically
that in shell models for three dimensional turbulence Response functions

are inertial range quantities. This is a strong indication that
the invariant measure describing the shell-velocity fluctuations is
characterized
by short range interactions between neighboring shells.\\
Response functions and generalized response-functions play an important
role in any diagrammatic approach for non-linear out-of-equilibrium
systems. In this work we have presented the first numerical attempt
to measure some of their properties in a systematic way. More work
is needed, both numerical and analytical, in order to better
understand the detailed structure of the invariant measure governing
F/R relations.  \\

We acknowledge useful discussions with G. Boffetta and  V. L'Vov. 
This work has been partially supported by
the EU under the Grant
No. HPRN-CT  2000-00162 ``Non Ideal Turbulence''. GL thanks the Department
of Physics of L'Aquila University for the kind hospitality.

\newpage

\newpage

\newpage

\newpage

%\begin{figure}
%\centerline{\psfig{file=fig1.tex.eps,width=6cm}}
%\centerline{a)}
%\includegraphics[width=0.75\textwidth]{pdf_norm.tex.eps}
%\centerline{b)}
%\includegraphics[width=0.75\textwidth]{pdf_norm_log.tex.eps}
%\caption{p.d.f of halving times for the shells 12 and 8,
% with the renormalization : $t'_{1/2}=\frac{\tau_{1/2}-
%<\tau_{1/2}>}{\sqrt{<\tau_{1/2}^2>-<\tau_{1/2}>^2}}$ and
% $pdf(t'_{1/2})=pdf(t_{1/2})*{\sqrt{<\tau_{1/2}^2>-<\tau_{1/2}>^2}}$}
%\label{pdfHT}
%\end{figure}

\begin{thebibliography}{99}
\bibitem{note1}
with ``statistical equilibrium'' we intend the statistical properties
 described by the  invariant probability measure
of the dynamical system.
\bibitem{note} even moments, being positive definite, cannot be
considered response-function. They posses the typical
exponential growth with a characteristic time given by the lyapunov exponent
and than, for times large enough, they saturate to the averaged
maximal distance in the attractor.
\end{thebibliography}
\end{document}